# Anomalous Gate-tunable Capacitance in Van der Waals Heterostructures

## Authors

Linshang Chen[1,2], Haoran Long[1,2], Heng Wu[1,2], Rui Mei[1,2], Zhengyu Su[1,2], Mengjie Feng[1], Jiang-Bin Wu[1], Kenji Watanabe[4,5], Takashi Taniguchi[4,5], Xuewei Cao[3], Zhongming Wei[1,2,*], Ping-Heng Tan[1,2,*] , Yanmeng Shi[1,2,*]

## Affiliations

[1]State Key Laboratory of Semiconductor Physics and Chip Technologies, Institute of Semiconductors, Chinese Academy of Sciences, Beijing 100083, China
[2]Center of Materials Science and Optoelectronics Engineering, University of Chinese Academy of Sciences, Beijing 100049, China
[3]School of Physics, Nankai University, Tianjin 300071, China
[4]Research Center for Functional Materials, National Institute for Materials Science, Tsukuba, Japan
[5]International Center for Material Nanoarchitectonics, National Institute for Materials Science, Tsukuba, Japan
*Corresponding author: E-mail: zmwei@semi.ac.cn; phtan@semi.ac.cn; ymshi@semi.ac.cn

## Abstract

**The ferroelectricity emerging in non-polar graphene/hexagonal boron nitride (hBN) heterostructures has drawn considerable attention because of its fascinating properties and promising high-frequency electrical polarization switching. Yet, the underlying mechanism is still under debate. Here in twisted double bilayer graphene (TDBLG) aligned with its neighboring hBN, we observed two types of hysteresis - delayed hysteresis in top gate induced by the anomalous screening, and advanced hysteresis in back gate caused by the anomalous gate-tunable capacitance. To investigate the role played by moiré potential in the anomalous hysteresis, we studied a moiréless graphene heterostructure as control experiment. Unexpectedly, we observed exactly the same phenomena in this control device. Our findings suggest that the anomalous ferroelectricity in graphene/hBN heterostructures may originate from the dielectric material hBN, calling for further structural investigations on hBN. The observation of gate-tunable capacitance provides more insights in the mysterious ferroelectricity in graphene/hBN heterostructures, and should enable new design of memory devices such as memcapacitor based on tunable capacitance.**

## MAIN TEXT

### Introduction

Two-dimensional (2D) ferroelectricity with spontaneous electric polarization is not only fundamentally interesting in physics, but also technologically important towards the goal of device miniaturization and the non-von Neumann computing architectures. Of particular interest is the interface engineered ferroelectricity in van der Waals heterostructures composed of non-polar 2D materials such as graphene and hBN, where the electrical polarization switching is induced by the interlayer sliding. Such sliding ferroelectricity has been predicted and verified in van der Waals bilayers of hBN, transition metal dichalcogenides and many other van der Waals materials(*1–10*). Recently, aligned graphene/hBN heterostructures(*11–14*) provide one of the rare systems where moiré ferroelectricity is driven by charge ordering instead of conventional ionic displacement. The electronic ferroelectricity is promising for high-frequency electrical polarization switching and neuromorphic transistors(*15*, *16*). Despite its fascinating properties and potential applications, the underlying mechanism of electronic ferroelectricity in aligned graphene/hBN heterostructures remains to be uncovered.

Here in this work, we study the electronic transport in a twisted double bilayer graphene (TDBLG) device, in which two bilayer graphene are stacked and rotated by a small angle with respect to each other(*17–19*), as well as a simple moiréless monolayer graphene(MLG) device. Out of our expectation, we observed the same hysteretic phenomena in these two systems. We find that there exist two different types of hysteresis with gate voltages, one of which is in delayed fashion and the other one is advanced. By studying the Hall effect, we discovered that the former is caused by the anomalous gate screening, similar to previous reports, while the latter is induced by the gate-tunable effective capacitance, which is first explicitly reported to the best of our knowledge. The same hysteretic behaviors emerging in both moiré and moiréless devices suggests that the mysterious ferroelectric hysteresis in graphene/hBN heterostructures may originate from the dielectric material hBN, calling for further investigations on the structure of hBN.

## Results

### Two types of hysteresis in TDBLG

We first show the results obtained from the TDBLG device. Moiré engineering is an effective approach to create novel quantum states arising from strong electronic correlations(*20*, *21*). TDBLG is a highly tunable correlated system with rich phase diagram(*17–19*, *22–25*). The correlated states are highly sensitive to twist angle, out-of-plane displacement field and carrier density, and are demonstrated to be spin-polarized and electrically switchable magnetic. The combination of magnetic states and interface engineered ferroelectricity in TDBLG is promising to realize multiferroicity (*26*, *27*). The stack was fabricated using standard cut-and-stack transfer method, and TDBLG was encapsulated by two hBN flakes. During the fabrication, TDBLG is intentionally aligned with its top neighbouring hBN, and the top and bottom hBN are rotated by ≈ 30°. A dual-gate configuration was used, allowing to tune charge carrier density $n$ and out-of-plane displacement field $D$ independently. The details of the sample fabrication can be found in Materials and Methods. Fig. 1A plots the schematic of the final device, and the determination of crystal orientation is shown in Fig. S1.

The basic characterization of the TDBLG is shown in Fig. S2, which plot four-terminal longitudinal resistance $R_{xx}$ ($V_{bg}$, $V_{tg}$), exhibiting typical transport features of TDBLG. In addition to the main charge neutrality point (CNP), two band insulating states appear at both electron and hole sides, resulting from the full fillings of 1$^{st}$ conduction and valence bands. Similar to previous reports, the increasing out-of-plane displacement field $D$ first closes and then reopens the gap at the main CNP. The charge carrier of the full filling of 1$^{st}$ band $n_0 \approx 6.9 \times 10^{12}$ cm$^{-2}$, corresponding to the twist angle 1.72° of TDBLG. The twist angle is also confirmed by the Brown-Zak oscillation as shown in Fig. S3. In addition to the typical transport properties of TDBLG, we also observed anomalous hysteretic behaviours. As shown in Fig. S2, there is a huge resistance hysteresis between forward and backward scans of $V_{bg}$. Usually the resistance peaks in dual-gate mapping run from top left to bottom right with a negative slope -$C_{bg}/C_{tg}$, where $C_{bg}$ and $C_{tg}$ are the capacitance per area of back and top gates, respectively. Therefore, the three resistance peaks should run parallel to each other. However, the slope of the resistance peak ridge at the full filling of 1$^{st}$ valence band is much larger than the slopes at main CNP and the full filling of 1st conduction band (Fig. S2A). If we reverse the scan direction, scanning $V_{bg}$ backward from positive side to negative side as shown in Fig. S2B, the band insulating state at the full filling of 1st valence band behaves normally while the slope of the insulating state at the full filling of 1st conduction band is larger than expected. We also measure the data by stepping $V_{tg}$ from -9 to +9V, shown in Fig. S2D-F, revealing that the stepping direction doesn't influence the transport behaviour.

**Layer-specific anomalous screening in top gate**

To investigate this anomalous hysteresis in detail, we first show the resistance modulation of top gate voltage $V_{tg}$ and back gate voltage $V_{bg}$, which exhibit asymmetric behaviours. Fig. 1B displays $R_{xx}$ ($V_{tg}$) when $V_{bg}$ = 0. The resistance peak is the location of the CNP, where the density of states (DOS) is at minimum. A huge ferroelectric hysteresis appears between the forward and backward scans. This hysteresis is similar to that observed in bilayer graphene(BLG)/hBN heterostructures(*11–13*), and it appears in a delayed fashion, that is, the resistance peak appears at positive $V_{tg}$ side for forward scan (blue), and negative $V_{tg}$ side for backward scan (orange). Fig. 1C plots $R_{xx}$ ($V_{bg}$), for both forward (blue) and backward (orange) scans when $V_{tg}$=0. Similar to the results as a function of $V_{tg}$ shown in Fig. 1B, a huge hysteresis is also present as a function of $V_{bg}$. On the other hand, however, there are two extraordinary different features by comparing Fig. 1B and Fig. 1C. First, the hysteresis in Fig. 1C manifests in an advanced fashion, i.e., the CNP resistance peak appears on negative $V_{bg}$ side for forward scan, and on positive side for backward scan, while it is in a delayed fashion in Fig. 1B. Second, the magnitudes of the resistance peaks in forward and backward scans in $V_{bg}$ are different dramatically. As shown in Fig. 1C, the resistance peak in backward scan is ≈ 4.7 kΩ, while it is ≈ 2.3 kΩ in forward scan. In Fig. 1B, the magnitudes of the CNP resistance peaks are nearly identical. The different resistance peak magnitudes suggest that the out-of-plane displacement field differs dramatically for forward and backward scans. Therefore, there exist two different types of hysteresis in our TDBLG device, and the device is gated asymmetrically by top and back gates. Similar anomalous hysteresis has been reported in previous works(*11*, *12*, *26*, *28*), where only one gate is hysteretic and the other gate is non-hysteretic and serves as a control gate. In our work, however, both gates are hysteretic, exhibiting two different types of hysteresis.

Now we have shown that this dual-gate TDBLG device is gated asymmetrically by top and back gate voltages. We first focus on the transport behaviour tuned by $V_{tg}$. Fig. 2A and B plot $R_{xx}$ ($V_{tg}$, $V_{bg}$) mappings for forward and backward scans of $V_{tg}$, respectively. $V_{bg}$ is slowly stepped from -80 V to 80V. The dual-gate mappings in Fig. 2A and B apparently can be separated into two different regions. In the forward scan mapping (Fig. 2A), the CNP resistance peak ridge runs parallel to $V_{tg}$ axis for the gate voltage from -9V to -5.2V, suggesting the top gate voltage is screened out and not working. Beyond this screening region, the CNP resistance peak ridge behaves normally and depends on both gates. Similarly, a screening region appears in the range 5.2V to 9V in the backward scan (Fig. 2B). Following previous reports(*11*, *12*), we refer this screening region as layer-specific-anomalous-screening (LSAS). Comparing Fig. 2A and B, we note that LSAS region only appears when we reverse the scan directions of $V_{tg}$. In the normal region, the slope of the resistance peak is ≈ -6.2, which is the coupling ratio of back and top gate voltages $-C_{tg}/C_{bg}$. Similar behaviours have been observed in the bilayer graphene moiré heterostructures aligned with hBN. Fig. 2C displays the resistance difference between Fig. 2A and B. A huge hysteresis loop is clearly observed and the formation of hysteresis can be regarded as by shifting the resistance peak ridges along $V_{tg}$.

To further study the gating of $V_{tg}$, we measured the Hall effect, $R_{xy}$ at $B = \pm 1$ T, and symmetrize $R_{xy}$ (1T) = ($R_{xy}$ (+1T)- $R_{xy}$ (-1T))/2. The calculated Hall density is shown in Fig. 2D. In the forward scan, the Hall density is almost independent on $V_{tg}$ in the LSAS region, and increases linearly afterwards. This behaviour is reversed in the backward scan, forming a hysteresis loop. By fitting $n_{Hall}$ ($V_{tg}$) in the linear region, we can extract the capacitance of top gate $C_{tg} \approx 5.3 \times 10^{-4}$ F/m$^2$. Moreover, we also observed the similar ratchet effect as reported in Ref(*13*), shown in Fig. S4A. Ferroelectric hysteresis usually depends on the scan limit and history of external electric field. In Fig. S4C-D, we checked the hysteresis by gradually increasing $V_{tg}$ scan limits from 0 to ±9V. When $V_{tg}$ scan range is within -1.9 to +1.9V, the system is trapped in the screening region. This is also confirmed by the dual gate mapping shown in Fig. S4E-F. In this measurement, we limit $V_{tg}$ between -1 and +1 V, and step $V_{bg}$ from -80V to +80V. Surprisingly, the hysteresis disappears and the CNP peak ridges only run parallel to $V_{tg}$ axis. When scan limit is larger than 1.9V, the hysteresis appears and positions of CNP in forward and backward scans are independent on scan limits.

**Hysteresis and Hall effect measurement in back gate**

We now turn to investigate the hysteresis as a function of back gate voltage. Fig. 3A and B display $R_{xx}$ ($V_{bg}$, $V_{tg}$) mappings by fast-scanning $V_{bg}$ and step-scanning $V_{tg}$, and the resistance difference is shown in Fig. 3C. It's the same data as Fig. S2, but with a smaller $V_{bg}$ scan range between -40V and +40V for clearer analysis. Similar to the observations in Fig. S2, we find that the CNP resistance peaks follow two different slopes. In Fig. 3A of forward scan, the slope for the range of -15.7V to 40V is ≈ -0.16 = - $C_{bg}$ / $C_{tg}$, while from $V_{bg}$ = -40V to -15.7V, the slope strikingly doubled to ≈ -0.32 = -2*$C_{bg}$ / $C_{tg}$. Similarly, in Fig. 3B of backward scan, the slope for the range of -40V to 15.7V is ≈ -0.16, and it is ≈ -0.32 from $V_{bg}$ = 15.7V to 40V. The region with the slope of - $C_{bg}$ / $C_{tg}$ ≈ -0.16 correspond to the normal state, where the change of charge carrier induced by one gate is compensated by the other one. Furthermore, when we reduce the scan range to even smaller between -10V and 10V, we find that the system is stuck in the anomalous slope region, shown in Fig. S5. In this mapping, we fast-scan $V_{bg}$ from -10V to 10V, and step $V_{tg}$ from -9V to 9V. The slopes of the CNP resistance peaks are both -0.32. Moreover, the hysteresis disappears when the system stuck in the anomalous slope region. The observation is very similar to that of $V_{tg}$, except that $V_{tg}$ is screened in small scanning range. This doubled slope is very bizarre and unexpected, suggesting that the effective capacitance of the back gate is twice as the normal capacitance. More data of the dual-gate mappings is shown in Fig. S6.

To further understand the two regions with different slopes, we measure the Hall effect as a function of $V_{bg}$ and extract the effective capacitances. First, we study the Hall effect when $V_{bg}$ scan limit is small, between -10V and 10V, where the system is trapped in the anomalous slope region, and the Hall carrier density for this region is displayed in Fig. 3D. Magenta dash line is the linear fit to $n_{Hall}$, and the fitting capacitance is $C_{effective} \approx 1.7 \times 10^{-4}$ F/m$^2$. In our device, we use heavily doped silicon substrate as back gate, and the dielectric material is 300nm-thick $SiO_2$, of which the geometry capacitance per unit area is $C_{geometry} = \varepsilon_r\varepsilon_0/d$, where $\varepsilon_r$ = 3.9 is the dielectric constant of $SiO_2$, $\varepsilon_0$ the vacuum permittivity and $d$ = 300 nm the thickness of $SiO_2$. The estimated back gate geometry capacitance of our sample should be $C_{geometry} \approx 1 \times 10^{-4}$ F/m$^2$. Therefore, the extracted $C_{effective}$ is much larger than expected, not to mention that the bottom hBN is on the order of tens of nanometers, so that the geometry capacitance should be further reduced. We subsequently increase the scan range of $V_{bg}$ to ±80V and repeat the Hall measurements. The results are shown in Fig. 3E for forward (blue) and backward (orange) scans, respectively. In both scans, the CNP resistance peaks locate in the normal TDBLG transport behaviour region, and the extracted back gate capacitance $C_{bg} \approx 0.85 \times 10^{-4}$ F/m$^2$, as denoted by the magenta dash lines. On the other hand, in the anomalous region, the effective back gate capacitance is doubled and estimated to be $\approx 1.7 \times 10^{-4}$ F/m$^2$, shown by the cyan dashed lines, which corresponds to the same value extracted in Fig. 3D.

With this observation of unexpected larger effective back gate capacitance, we are able to explain the aforementioned two unexpected different features between Fig. 1B and 1C. First, for an undoped graphene device, the CNP resistance peak should locate at zero gate voltage. But if the effective capacitance is doubled when we start scanning and then goes back to normal, graphene will behave like *n*-doped (*p*-doped) for forward (backward) scan at $V_{bg} = 0$, that is, the CNP resistance moves to negative (positive) gate voltage in forward (backward) scan, giving rise to the advanced hysteresis. Second, although the effective capacitance doubles, only half of the charges are injected from ground to the bottom graphene layer through geometry capacitance, and the other half should originate from some other mechanism. Considering that the resistance peak magnitude varies dramatically between forward and backward scans, we speculate that the other half of charges are injected to the top graphene layer, causing a finite displacement field difference between forward and backward scans. Therefore, electrons are injected to top graphene layer in the forward scan, and removed from the top graphene layer in the backward scan. In this case, the effective back gate capacitance is doubled, but $D$ are different for forward and back scans, giving rise to different band gaps and resistance peak magnitudes.

So far, we have established the fact that the top gate is anomalously screened, while the back gate is anomalously more efficient in our TDBLG device. Thanks to the dual-gate configuration of our sample, we can tune the external carrier density $n_{ext}$ and external displacement field $D_{ext}$ independently by calculating $n_{ext} = (C_{bg}V_{bg} + C_{tg}V_{tg})/e$ and $D_{ext} = (C_{bg}V_{bg} - C_{tg}V_{tg})/2\varepsilon_0$ where $\varepsilon_0$ is the vacuum permittivity. When we fix $D_{ext}$ and scan $n_{ext}$, the hysteretic behaviour is absent, displayed in mapping $R_{xx}(n_{ext}, D_{ext})$ in Fig. S7. Fig. 3F displays a line cut of $R_{xx}$ as a function $n_{ext}$ under $D_{ext} = 0.1$ V/nm. However, we note that the CNP peak is located at positive $n_{ext}$ side, suggesting the graphene is *p*-doped. In this measurement, we scan $n_{ext}$ from hole to electron side, and step $D_{ext}$ from positive to negative, as indicated by the black arrows (Fig. S7A). Fig. S7B shows the same measurement, but we step $D_{ext}$ from negative to positive side. This time, the CNP peak moves to negative $n_{ext}$, suggesting graphene is *n*-doped. Finally, when we fix $n_{ext}$ and scan $D_{ext}$, shown in Fig. 3G and Fig. S8, the huge hysteresis emerges. We also note the modulation of $D_{ext}$ seems changing graphene carrier density, which is dramatically different from normal graphene devices.

Antiparallel (AB-BA) stacked TDBLG has been predicted to host ferroelectricity due to interlayer sliding(*29*–*31*). However, this case doesn't apply to our AB-AB stacked sample. The external electric field needed to flip the sliding polarization is less than 0.05V/nm(*29*), one order of magnitude smaller than $\approx$0.5V/nm in our case (Fig. 3G). The LSAS in graphene/hBN heterostructures has been attributed to the moiré potential(*12*, *13*) or the defects in hBN(*28*, *32*). The observation of anomalous gate-tunable capacitance represents a novel finding. The most straightforward way to determine the role of moiré potential is to check if the same phenomena exist in moiréless control devices.

## Hysteresis in moiréless MLG device

To this end, we fabricated a simpler moiréless device, monolayer graphene (MLG) without any moiré involved. Fig. 4A displays the device structure and optical microscope picture. MLG is encapsulated between top and bottom hBN flakes, and the stack was deposited on $SiO_2$/Si substrate. Fig. 4B and C display the four-terminal resistance as a function of $V_{tg}$ and $V_{bg}$, respectively. Surprisingly, two types of hysteresis, similar to those in the TDBLG device (Fig. 1B–C), also appear in the moiréless MLG device. The hysteresis as a function of $V_{tg}$ is in delayed fashion (Fig. 4B), where the CNP appears on positive $V_{tg}$ in forward scan and on negative $V_{tg}$ in backward scan. On the contrary, as shown in Fig. 4C, the CNP appears on negative $V_{bg}$ in forward scan and on positive $V_{bg}$ in backward scan, manifesting an advanced hysteresis as a function of $V_{bg}$.

As demonstrated in TDBLG device, the delayed hysteresis originates from the anomalous screening of top gate, and the advanced hysteresis is induced by the anomalous gate-tunable capacitance of back gate. We also measured the Hall effect and studied the Hall carrier density in the MLG device, shown in Fig. 4D-E. Indeed, the modulation of the carrier density by gate voltages in the MLG device exhibits exactly the same behaviour as in the TDBLG device. As displayed in Fig. 4D, in the forward scan of $V_{tg}$ (blue), the carrier density is independent on gate voltage first between -6 V and -2.2 V, and then linearly increases with $V_{tg}$ as a normal MLG device. In the backward scan (orange), $n_{Hall}$ is also independent on $V_{tg}$ first between 6 V and 2.2 V, then behaves normally. The top gate capacitance fitted from the linear region is $C_{tg} \approx 3.6 \times 10^{-4}$ F/m$^2$ (magenta dash line). Plotted in Fig. 4E, $n_{Hall}$ ($V_{bg}$) can also be separated into two regions with two different slopes (effective capacitance). In the backward scan of $V_{bg}$ (orange), the slope of the region between 5V and 18V with an effective capacitance of $C_{effective} \approx 2 \times 10^{-4}$ F/m$^2$ (cyan dash line) is much larger than that from 5V to -18V with an effective capacitance of $C_{effective} \approx 0.68 \times 10^{-4}$ F/m$^2$ (magenta dash line). Again, $C_{effective} \approx 2 \times 10^{-4}$ F/m$^2$ is much larger than that expected from the geometry capacitance of 300 nm thick $SiO_2$ and tens of nanometer thick hBN flake. In contrast to a factor of two between the anomalous effective capacitance and geometry capacitance in TDBLG, the anomalous capacitance is 2.94 times as large as the geometry one in the MLG device. The order of the effective capacitance is reversed in the forward scan (blue), giving rise to a hysteretic loop. More characterisation of these two types of hysteresis in MLG is shown in Fig. S9.

Fig. 4F presents the dependence of the resistance on the external carrier density (left) and external displacement field (right). A minor hysteresis appears in $R_{xx}$ ($n_{ext}$) when $D_{ext}$ = 0, and it can be diminished by the application of external $D_{ext}$ = 0.33 V·nm$^{-1}$. This observation is consistent with the fact that the polarisation is tuned by the external displacement field. $R_{xx}$ ($D_{ext}$) displays huge hysteresis, similar as the TDBLG device. The mappings of $R_{xx}$ ($n_{ext}$, $D_{ext}$) and $R_{xx}$ ($D_{ext}$, $n_{ext}$) are plotted in Fig. S10 and S11.

So far, there are two different scenarios trying to explain the LSAS in graphene/hBN heterostructures. The first one involves the interplay between the itinerant charges in graphene and the localised charges caused by the interfacial moiré potential(*12*, *13*). The other explanation points to the charge redistribution in dielectric hBN(*28*), which could be caused by stacking faults or defects. The ideal layered hBN with AA' structure is non-ferroelectric. The stacking faults(*32*) or rhombohedral stacking(*2*)(*33*) could give rise to a ferroelectric polarization within hBN flake. Since we observed the same anomalous gate-screening and gate-tunable capacitance in both moiré and moiréless devices, the former explanation can be ruled out in our work, as there is no moiré potential involved in our MLG device. Therefore, the phenomena reported in this work should originate from the properties of the dielectric materials hBN. However, both scenarios are utilized to explain the observation of anomalous screening. The observation of gate-tunable capacitance in our work is new, hinting the existence of some new defects' configurations. Further detailed theoretical and experimental investigations on the structure of hBN flakes are required to nail down the underlying mechanism.

Although we don't have a satisfactory microscopic theory on the mechanism, here we propose a phenomenological model with an effective electrical polarization, which produces an out-of-plane displacement field $D_p$ that can be tuned by external $D_{ext}$, based on the observations of the asymmetric gating in our sample, i.e., the top gate is anomalously screened, while back gate is anomalously more efficient in injecting charges into graphene, The schematic of this model is plotted in Fig. S12. Without claiming the nature of the polarization, which could be caused by the stacking faults or charge redistribution within hBN defects, we proposed that the effective polarization is located between top gate and graphene. Therefore, the total charge carrier in graphene is $n_{tot} = n_t + n_p + n_b = \frac{2\varepsilon_0}{e}(-D_t - D_p + D_b)$, where $n_t$, $n_p$, $n_b$ are chargers induced by top gate displacement field $D_t$, the polarization displacement field $D_p$ and back gate displacement field $D_b$, respectively. Here we assume the change of $D_p$ is proportional to $D_t$, and $D_b$ , $\Delta D_p = \alpha\Delta D_t + \beta\Delta D_b$, and $D_p$ saturates at $\pm D_{saturation}$, so the $D_p$ could have different dependence on $D_t$ and $D_b$. When we scan $V_{tg}$, $\Delta D_p = \alpha\Delta D_t$, so the change of total carrier density $\Delta n_{tot} = (1+\alpha)\Delta n_t$. To explain the screening of top gate, we attribute $\alpha$ to be -1. In this case, $n_{tot}$ is independent on $V_{tg}$. When $D_p$ is saturated, $n_{tot}$ is only tuned by top gate and the system goes into normal region. When we scan $V_{bg}$, $\Delta D_p = \beta\Delta D_b$, so $\Delta n_p = \beta\Delta n_b$ and $\Delta n_{tot} = (1-\beta)\Delta n_b$. We attribute $\beta$ = -1 for the TDBLG device. In this case, $V_{bg}$ is doubly effective in injecting electrons or holes into graphene. However, $\Delta n_p$ accumulates on the top graphene layer, and $\Delta n_b$ accumulates on the bottom graphene layer. Therefore, the total displacement field is different for forward and backward scans of $V_{bg}$. When $D_p$ is saturated, $n_{tot}$ is only tuned by back gate and the system goes into normal region. For the moiréless MLG, we attribute $\beta$ = -1.94, so that the anomalous effective $C_{bg}$ is 2.94 times as large as its geometry capacitance.

## Discussion

Now we provide a brief discussion on the plausible nature of the polarization. One key question is whether moiré pattern at interface is a critical factor. Although there are evidences suggesting that an effective polarization could exist resulting from the interplay between itinerant and moiré localised charges(*13*)(*15*), we can preclude this possibility in our MLG device as no moiré pattern is involved. Another possibility of sliding ferroelectricity within hBN layers is also very unlikely since the required switching electric field is one order of magnitude smaller than our observation. Therefore, there could exist an unidentified non-ideal AA' stacking structure in hBN flakes that gives rise to an effective polarisation field, such as the proposed conducting channels in previous work(*32*). We note that a solid-state memcapacitor model with layered metamaterials consisting of alternating conducting and insulating layers has been proposed, that could potentially apply to the insulating hBN flakes with conducting layers(*34*). The tunnelling between conducting layers could give rise to hysteretic charge-voltage and capacitance-voltage curves. If future theory and experiment could identify the existence of the local conducting layers in hBN, new memory devices could be intentionally engineered and designed.

In summary, in a TDBLG moiré heterostructure and a MLG device without any moiré involved, we observed exactly the same hysteretic phenomena. First, two different types of ferroelectric hysteresis emerge in both devices. The delayed hysteresis is induced by the layer specific anomalous screening, while the advanced one is caused by the novel observation of gate-tunable effective capacitance. The same observation in both moiré and moiréless devices provides a strong evidence to rule out the role of moiré potential in interpreting the mysterious ferroelectric hysteresis in graphene/hBN heterostructures. Our work points out that more theoretical and experimental investigations should focus on the structural analysis of dielectric hBN. The history-dependence in effective capacitance, or memcapacitive behaviour could lead to the invention of graphene/hBN heterostructure based memory devices for neuromorphic circuits.

## Materials and Methods

### Sample fabrication and electrical transport measurement

The graphene and hBN nanosheets are mechanically exfoliated on a 285nm thick $SiO_2$/Si substrate, and the layer numbers is determined by optical microscope contrast and Raman spectroscopy. We use the stamp consisting of glass slide, PDMS dome, PPC film (15%) to assemble the heterostructure using the dry transfer technique. The stamp is employed to pick up the bottom hBN layer, and the graphene layer and hBN flakes can be picked up through van der Waals (vdW) force. For TDBLG, we cut the BLG using a tungsten tip. The alignment between graphene and hBN is achieved by aligning the sharp straight edges. On the other hand, for moiréless MLG, we intentionally misaligned the straight edges of MLG with that of top and bottom hBN flakes by 10° and 20°, respectively, to avoid the formation of long-range moiré potential. The moiréless configuration is further confirmed by transport measurements. The final stack is placed on a 285nm thick $SiO_2$/Si substrate, followed by dissolving PPC film in chloroform and annealing to remove the polymer residue. We follow the standard nanofabrication process to make devices from the stack. We define the electrodes with e-beam lithography and develop with chilled DI water/IPA (1:3), followed by the deposition of metal (5nm Cr and 50nm Au). We use the same process to define the top gate, and etch the samples into Hall bar using RIE or ICP. The doped Si substrate serves as back gate.

Resistance measurements were carried out using standard low-frequency lock-technique with excitation current of few nA range in a commercial cryostat equipped with superconducting magnet. All measurements were taken at the temperature of 1.5K unless specified otherwise. Gate voltages were applied by commercial DC voltage source meter units.

### Determination of crystal orientation of hBN

The second harmonic generation (SHG) technique was used to characterize the crystal orientation of hBN. In SHG measurements, a Ti:Sapphire femtosecond pulsed laser operating at a wavelength of 800 nm serves as the incident light with excitation power of a few milliwatts. The linearly polarized incident laser beam passes through a sequence of optical elements—half-wave plate, polarizer, analyser, and another half-wave plate—before reaching the surface of hBN. As the polarization rotates over $2\theta$ when the half-wave plate changes $\theta$, a 180° rotation is sufficient to capture the full dataset. The SHG signal is minimized when the polarization of incident light is along the zigzag direction, and maximized along the armchair direction. When plotting the SHG signal intensity against the polarization direction in a polar coordinate system, a sixfold polarized pattern with a period of 60° is anticipated. Fitting the pattern using $I_{SHG} \propto \cos^2(\theta_{pol} - \varphi)$, where the parameter $\varphi$ denotes the crystal orientation of hBN.

## Acknowledgments

The authors would like to thank K. S. Novoselov and D. Zhang for fruitful discussions.

**Funding:** This work was sponsored by the National Key Research and Development Program of China (Grant No. 2024YFA1409700), the National Natural Science Foundation of China (Grant No. 12274402, 62125404, U24A20285), Beijing Natural Science Foundation (Grant No. Z220005), the Science Foundation of the Chinese Academy of Sciences (Grant No. JCPYJJ-22012). K.W. and T.T. acknowledge support from the JSPS KAKENHI (Grant Numbers 21H05233 and 23H02052) and World Premier International Research Center Initiative (WPI), MEXT, Japan.

**Author contributions:**

Y.S. conceived the presented idea and directed the project. Y.S., L.C., H.L. and Z.W. performed transport measurements. Y.S., L.C., H.W., R.M., Z.S. and M.F. fabricated devices. Y.S., L.C. and H.L. performed data analysis. Y.S. and L.C. contributed to the interpretation of data. K.W. and T.T. grew hBN single crystals. Y.S. and L.C. contributed to the writing of the manuscript. All authors discussed the results and commented on the manuscript.

**Competing interests:** The authors declare no competing interests.

**Data and materials availability:** All relevant data are available from the corresponding author on reasonable request.

**Figures and Tables**

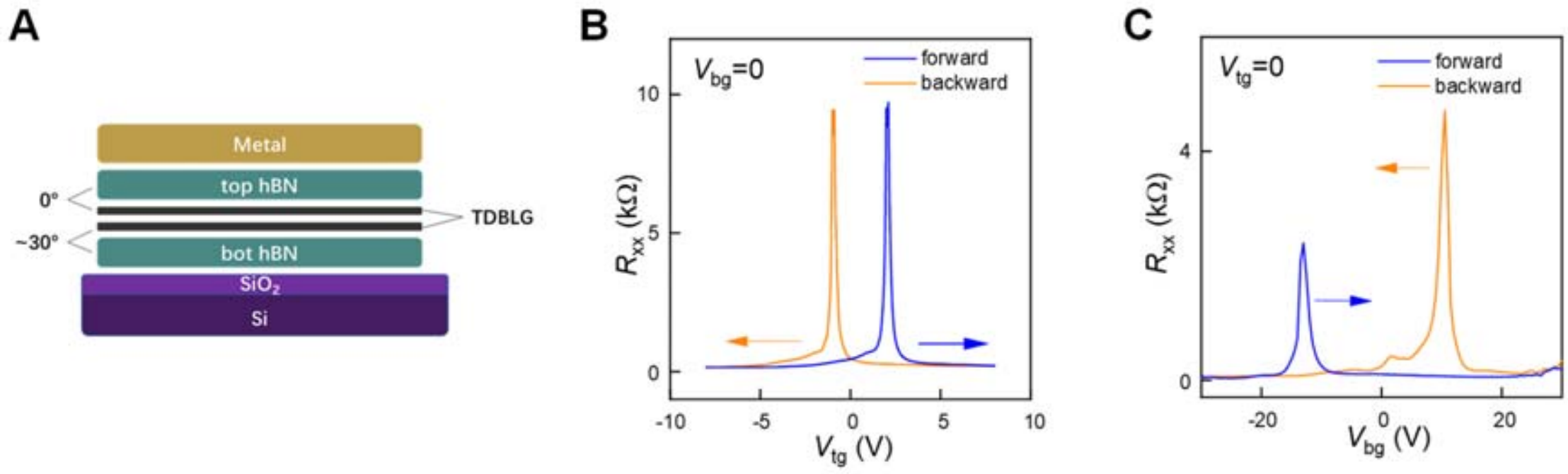

**Fig. 1 Two types of hysteresis in TDBLG device.** (**A**) Schematic of dual-gate TDBLG device. (**B**) Four-terminal longitudinal resistance $R_{xx}$ as a function of top gate voltage $V_{tg}$, for forward (blue) and backward (orange) scans, respectively, when $V_{bg} = 0$. (**C**) Four-terminal longitudinal resistance as a function of back gate voltage $V_{bg}$, for forward (blue) and backward (orange) scans, respectively when $V_{tg} = 0$.

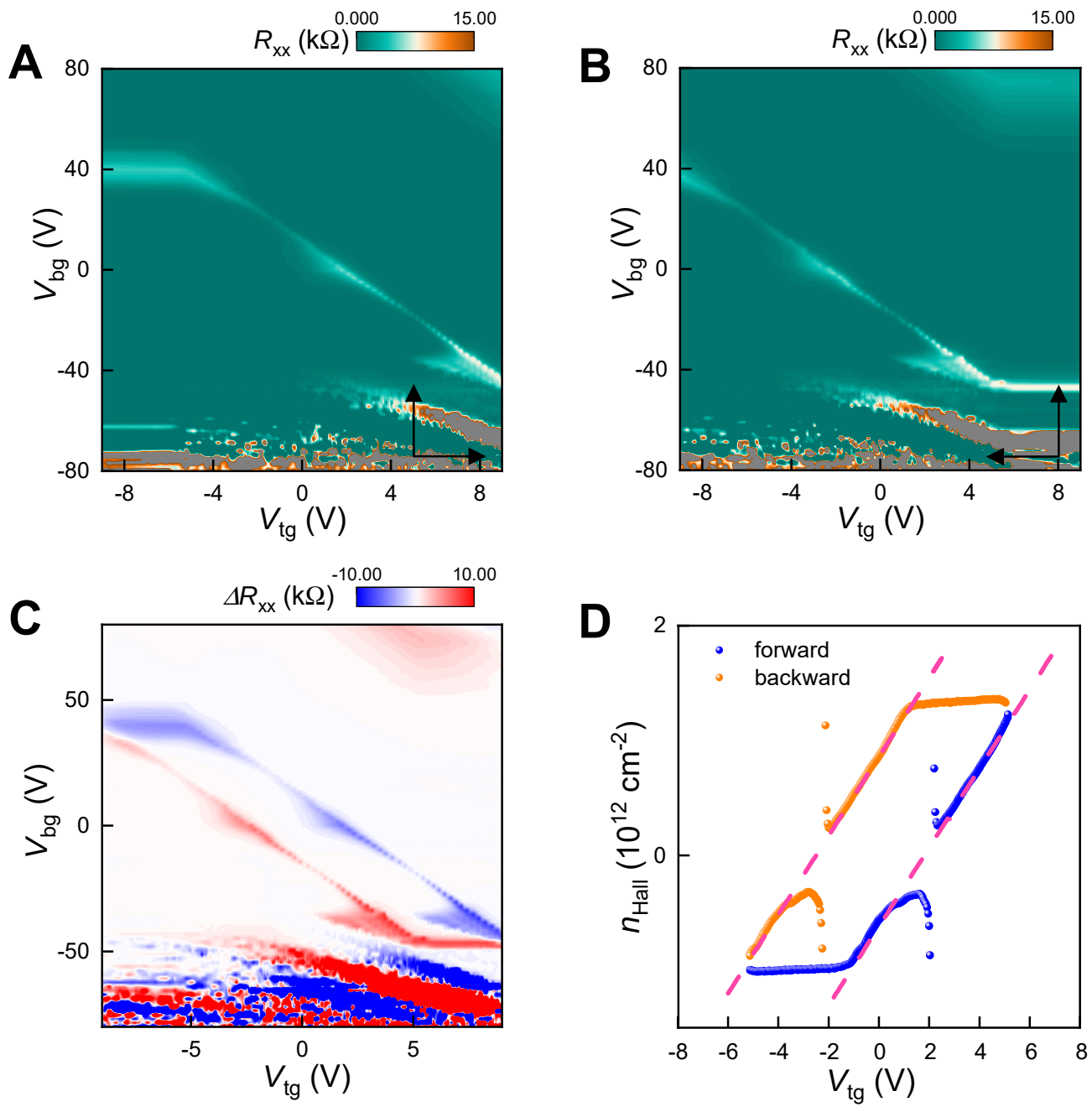

**Fig. 2 Layer-specific anomalous screening and hysteretic transport behaviour as a function of top gate voltage.** (**A-B**) Four-terminal longitudinal resistance mappings $R_{xx}(V_{tg}, V_{bg})$ for forward (A) and backward scans (B) of $V_{tg}$. Black arrows indicate the fast-scan directions of $V_{tg}$ and stepping direction of $V_{bg}$. (**C**) The resistance difference between (A) and (B). (**D**) Carrier density extracted from Hall effect measurement as a function of $V_{tg}$ when $V_{bg} = 0$. The fitted top gate capacitance $C_{tg}$ in normal regime is $\approx 5.3 \times 10^{-4}$ F/m$^2$.

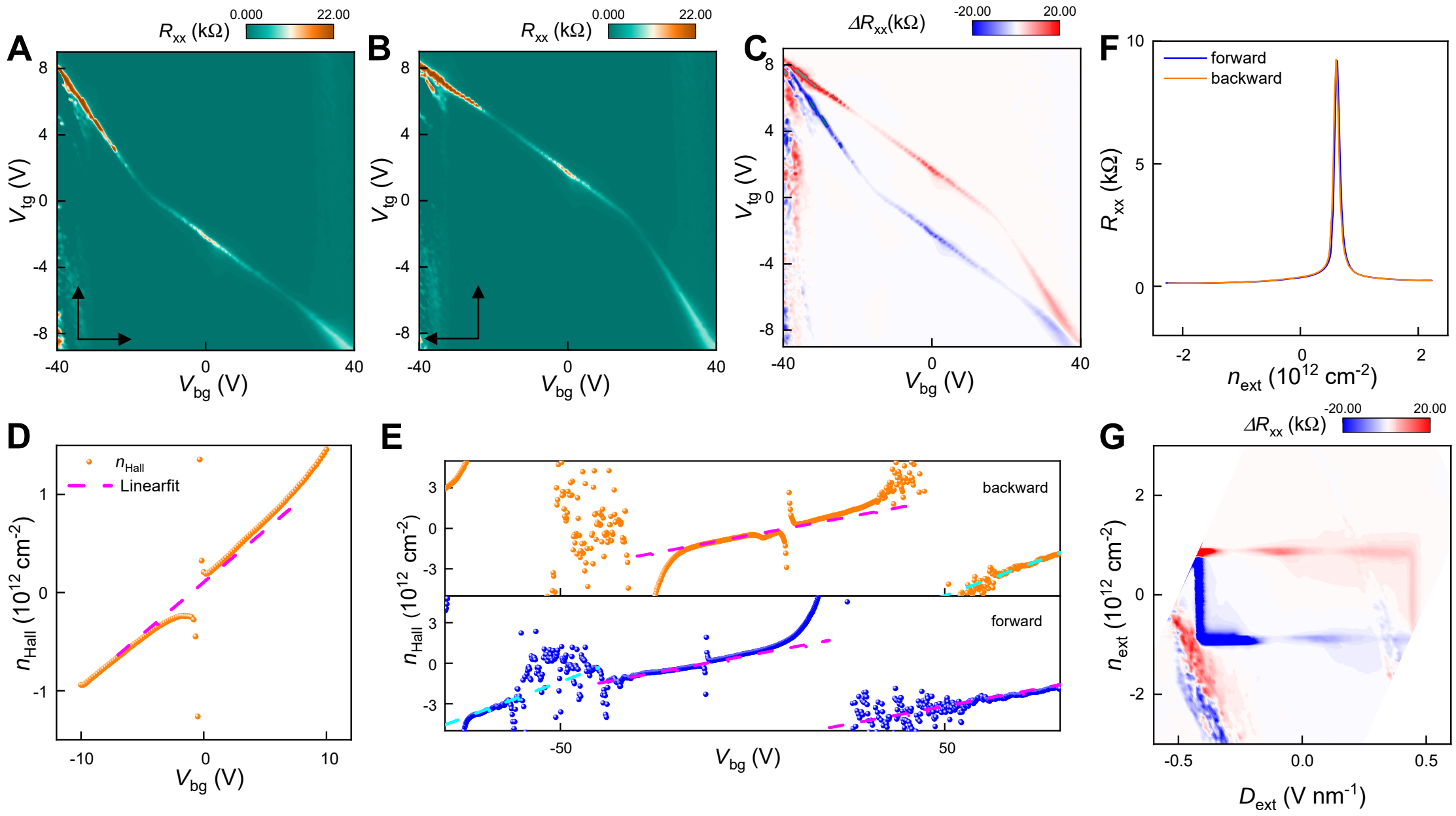

**Fig. 3 The anomalous gating and hysteretic transport behaviour as a function of back gate voltage.** (**A-B**) Dual-gate mapping of four-terminal longitudinal resistance $R_{xx}$ *(*$V_{bg}$, $V_{tg}$*)* for forward (A) and backward (B) scans of $V_{bg}$. The black arrows indicate the sweeping and stepping directions of gate voltages. (**C**) The resistance difference between (A) and (B)**. (D-E)** Hall carrier density as a function of $V_{bg}$ when $V_{bg}$ sweep limit is between (-10V, +10V) and the system is stuck in the anomalous region (D), and when $V_{bg}$ sweep limit is between (-80V, +80V) (E). Scattered lines are experimental data, and dash lines are linear fittings. The fitted effective capacitance $C_{effective}$ in (D) is ≈ 1.7 × $10^{-4}$ F/m$^2$ (magenta dash line). The fitted effective capacitance $C_{effective}$ in (E) is ≈ 1.7 × $10^{-4}$ F/m$^2$ (cyan dash lines) and ≈ 0.85 × $10^{-4}$ F/m$^2$ (magenta dash lines). (**F**) $R_{xx}$ ($n_{ext}$) when $D_{ext}$ = 0.1 V/nm, showing no hysteresis. (**G**) The resistance difference mapping $\Delta R_{xx}$ ($D_{ext}$, $n_{ext}$) between forward and backward scans of $D_{ext}$.

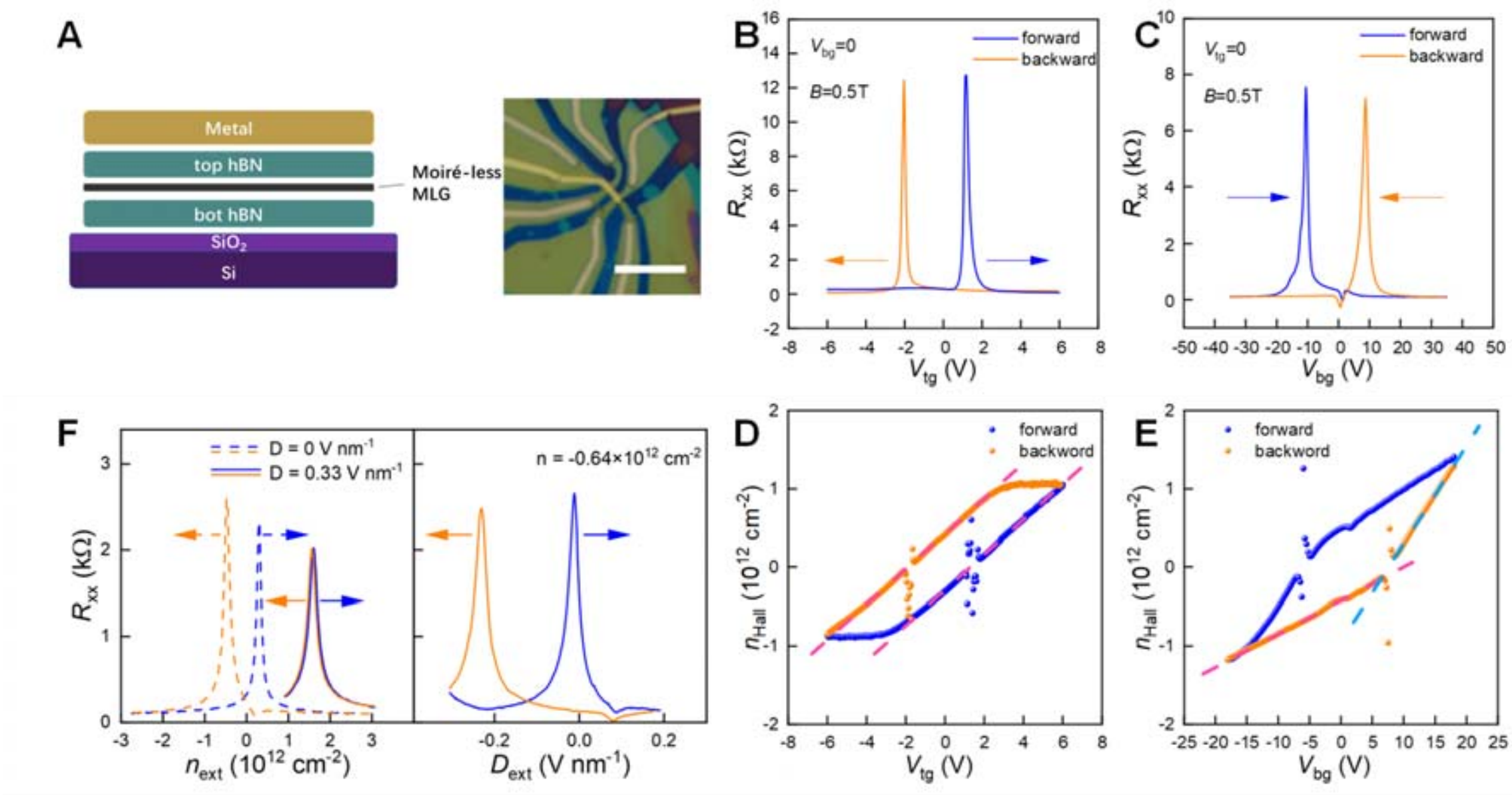

**Fig. 4 Anomalous screening and gate-tunable capacitance in a moiréless MLG device.** (**A**) Schematic (left) and the optical microscope picture of the moiréless MLG device structure. Scale bar is 10 µm. (**B-C**) The different hysteretic behaviours as a function of $V_{tg}$ (B) and $V_{bg}$ (C). (**D-E**) The dependence of $n_{Hall}$ on $V_{tg}$ (D) and $V_{bg}$ (E). The fitted top gate capacitance in (D) is $C_{tg} \approx 3.6 \times 10^{-4}$ F/m$^2$ (magenta dash line). The fitted effective capacitances of back gate in (E) are $C_{effective} \approx 2 \times 10^{-4}$ F/m$^2$ (cyan dash line) and $\approx 0.68 \times 10^{-4}$ F/m$^2$ (magenta dash line). (**F**) Line traces of $R_{xx}$ ($n_{ext}$) at $D_{ext}$ = 0 and 0.33 V·nm$^{-1}$ (left) $R_{xx}$ ($D_{ext}$) (right). The data is taken at $T$ = 200 K.

# Supplementary Materials for

## Anomalous Gate-tunable Capacitance in Van der Waals Heterostructures

Linshang Chen *et al.*

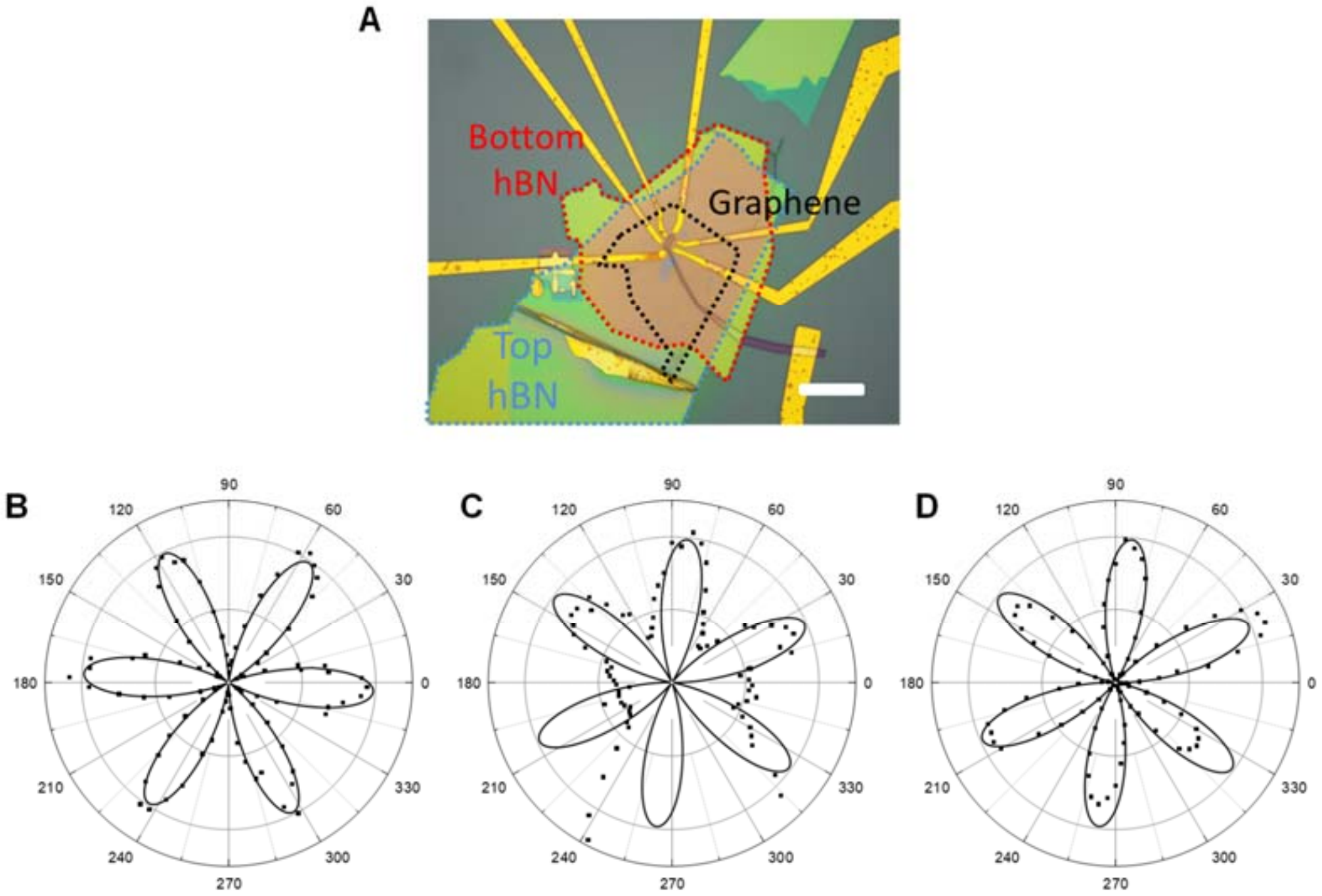

**Fig. S1. Alignment of TDBLG.** (**A**) Optical image of the encapsulated stack of TDBLG. Top hBN, TDBLG and bottom hBN are outlined by blue, black and red dot lines, respectively. Scale bar: 10 μm. (**B-D**) Second harmonic generation measurement of top hBN, the etched bottom hBN and a hBN flake, which locates near the bottom hBN and shares the same crystal orientation. The relative rotation angle between top and bottom hBN is ≈ 27°.

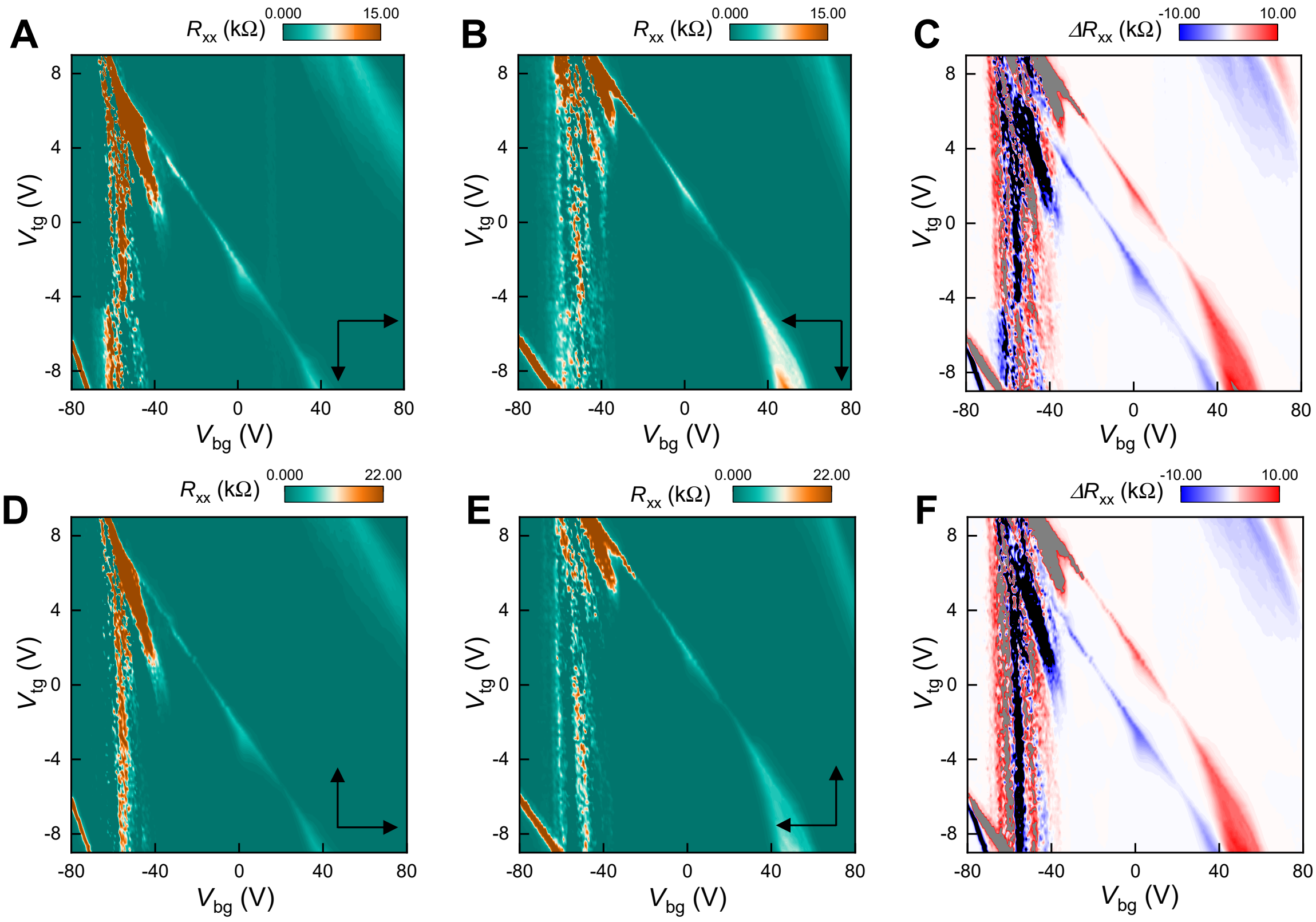

**Fig. S2. Trasport behaviour of TDBLG device by fast-sweeping $V_{bg}$.** (**A-B**) Dual-gate mapping of four-terminal longitudinal resistance $R_{xx}$ $(V_{bg}, V_{tg})$ for forward (A) and backward (B) fast-scans of $V_{bg}$. $V_{tg}$ was step-scanned from positive to negative. (**C**) The resistance difference between (A) and (B)**.** (**D-F**), Same data as (A-C), but with opposite stepping direction of $V_{tg}$. The black arrows indicate the sweeping and stepping directions of gate voltages.

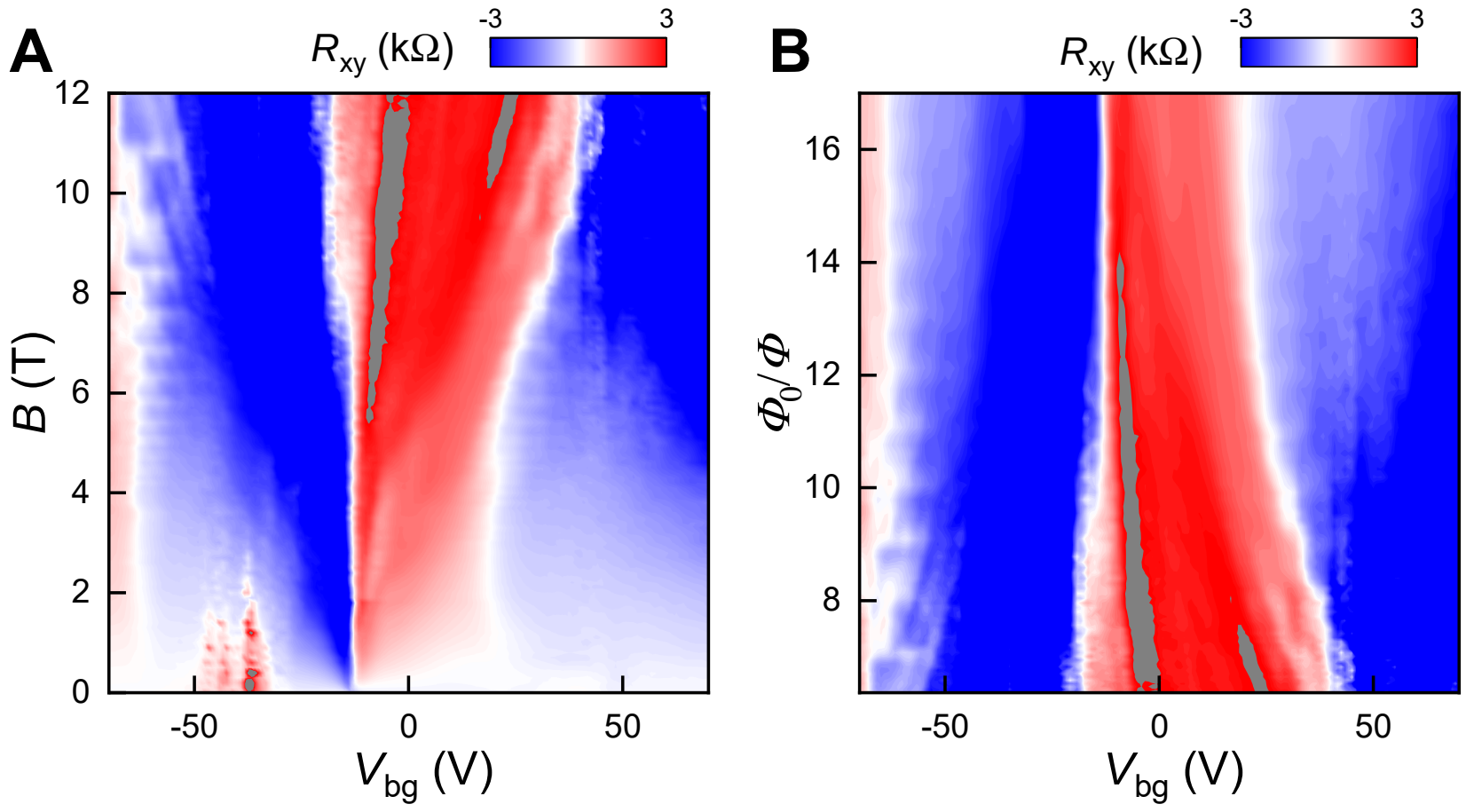

**Fig. S3. Transport behaviour of TDBLG device in perpendicular magnetic field.** (**A**) $R_{xx}(V_{bg}, B)$ mapping for backward scan of $V_{bg}$ when $V_{bg} = 0$. (**B**) Converted $R_{xx}(\Phi_0/\Phi\ , B)$ from (A), where $\Phi_0 = h/e$, $h$ is the Planck constant, $e$ is the electron charge, and $\Phi = BS$ is the magnetic flux through a moiré unit cell.

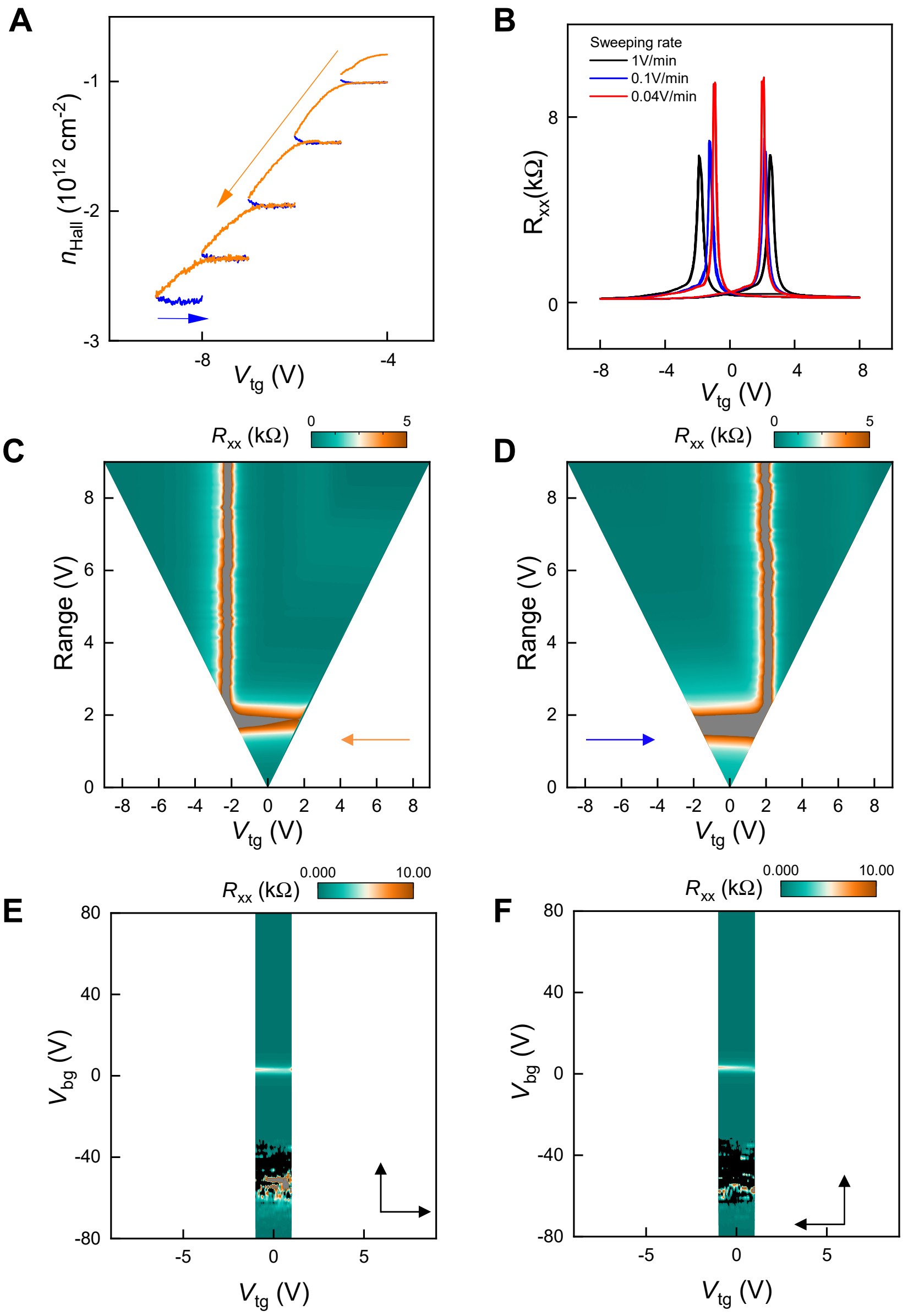

**Fig. S4. More characterization of the hysteresis as a function of $V_{tg}$ in TDBLG device.** (**A**) Ratchet effect on the hole side as a function of $V_{tg}$, $V_{bg}$=0. Orange and blue lines are backward and forward $V_{tg}$ scans, respectively. (**B**) Hysteresis as a function of $V_{tg}$ with sweeping rate of 1V/min (black), 0.1V/nm(blue) and 0.04V/min(red). The hysteresis is still very robust for a scan as long as 13.3 hours. (**C-D**) Hysteretic transport behaviour by gradually increasing $V_{tg}$ sweep limits for backward (C) and forward (D) scans. (**E-F**) $R_{xx}$ ($V_{tg}$, $V_{bg}$) dual gate mappings when $V_{tg}$ scan limit is between (-1V, 1V) for forward (E) and backward (F) scans.

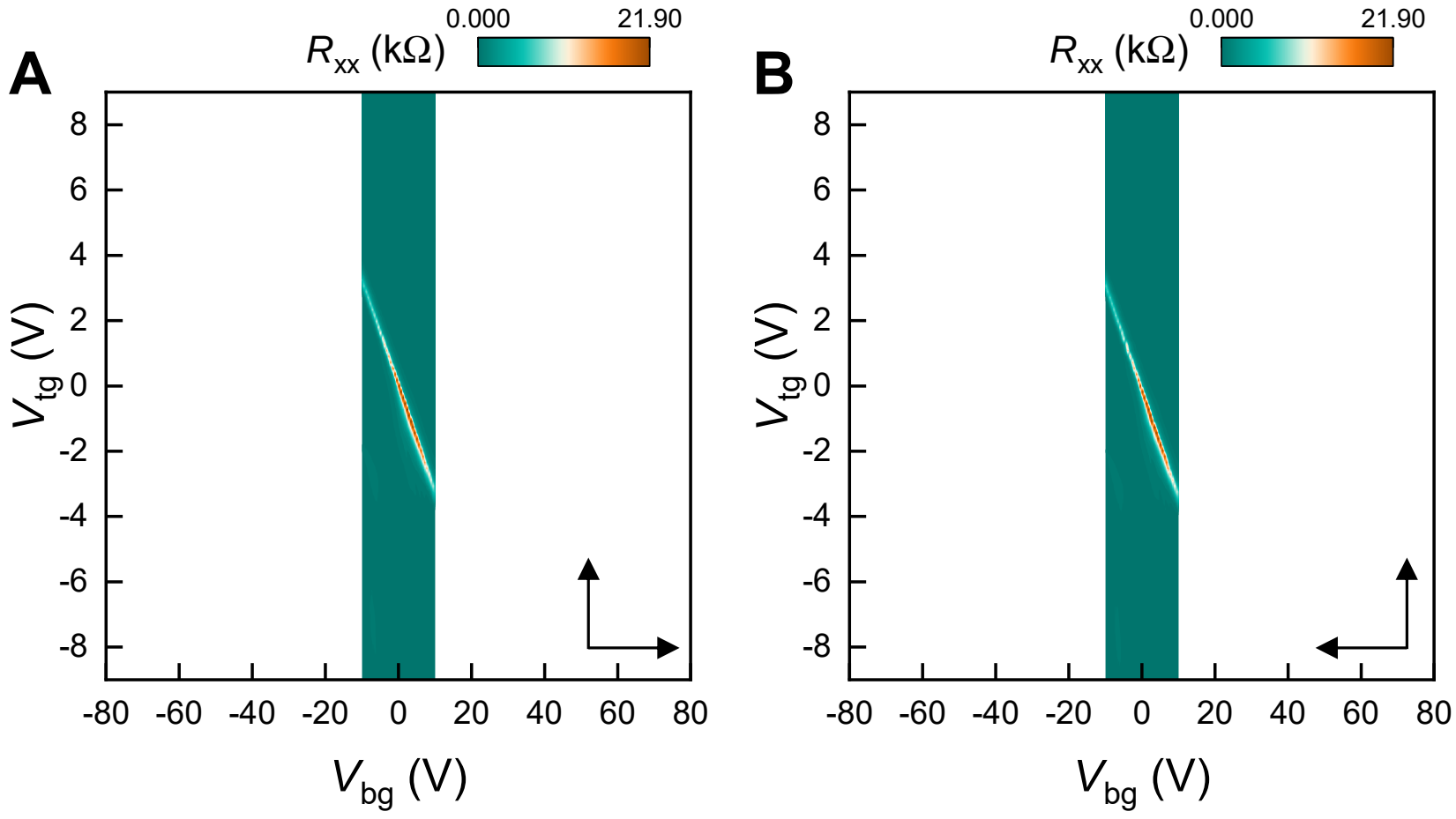

**Fig. S5. Anomalous gating of $V_{bg}$ in TDBLG device.** (**A-B**) $R_{xx}$ ($V_{bg}$, $V_{tg}$) dual gate mappings when $V_{bg}$ scan limit is between (-10V, 10V) for forward (A) and backward (B) scans.

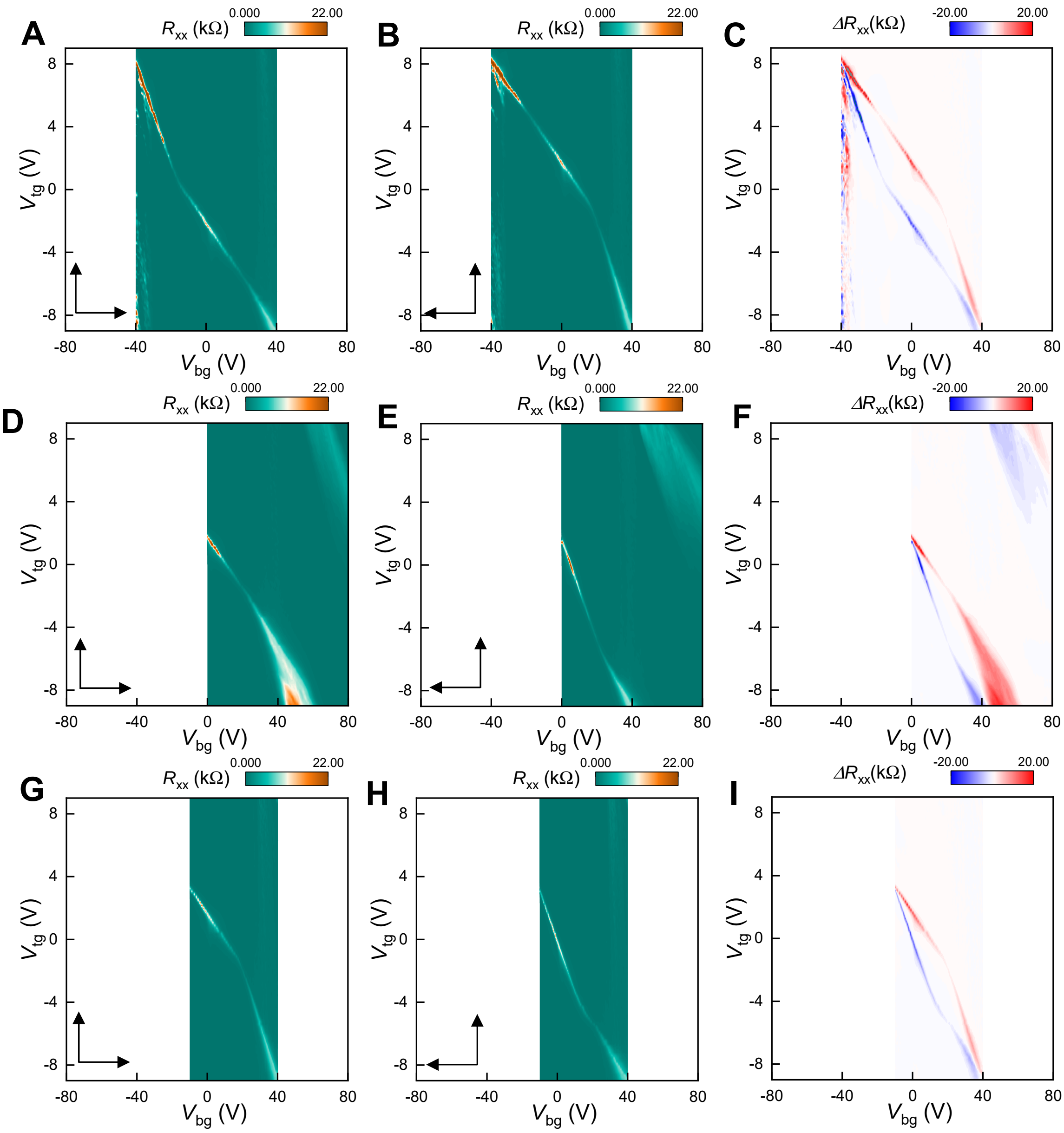

**Fig. S6. Additional data on the anomalous gating of $V_{bg}$ in TDBLG device.** $R_{xx}$ ($V_{bg}$, $V_{tg}$) dual gate mappings when $V_{bg}$ scan limit is within -40 to 40V (**A-C**), within 0 to 80V (**D-F**) and within -10 to 40V (**G-I**).

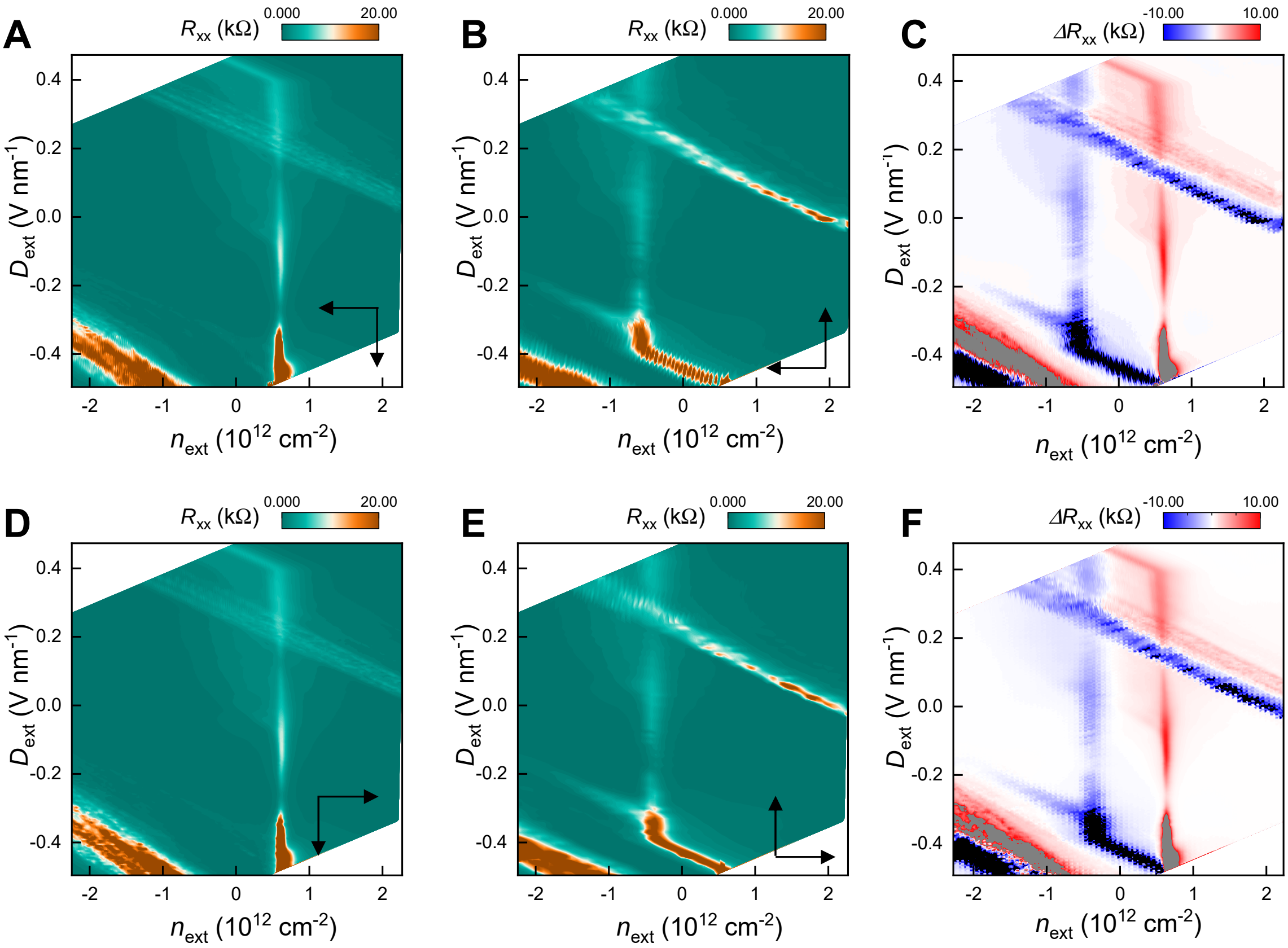

**Fig. S7. Additional data on the transport behaviour of TDBLG device when fast-scanning $n_{ext}$.** (**A-B**) Four-terminal longitudinal resistance $R_{xx}$ $(n_{ext}, D_{ext})$ for backward fast-scans of $n_{ext}$. $D_{ext}$ was step-scanned from positive to negative (A) and from negative to positive (B). (**C**) The resistance difference between (A)and (B). (**D-F**) Same data as (A-C), but with opposite fast-scanning direction of $n_{ext}$. The black arrows indicate the sweeping and stepping directions of $n_{ext}$ and $D_{ext}$.

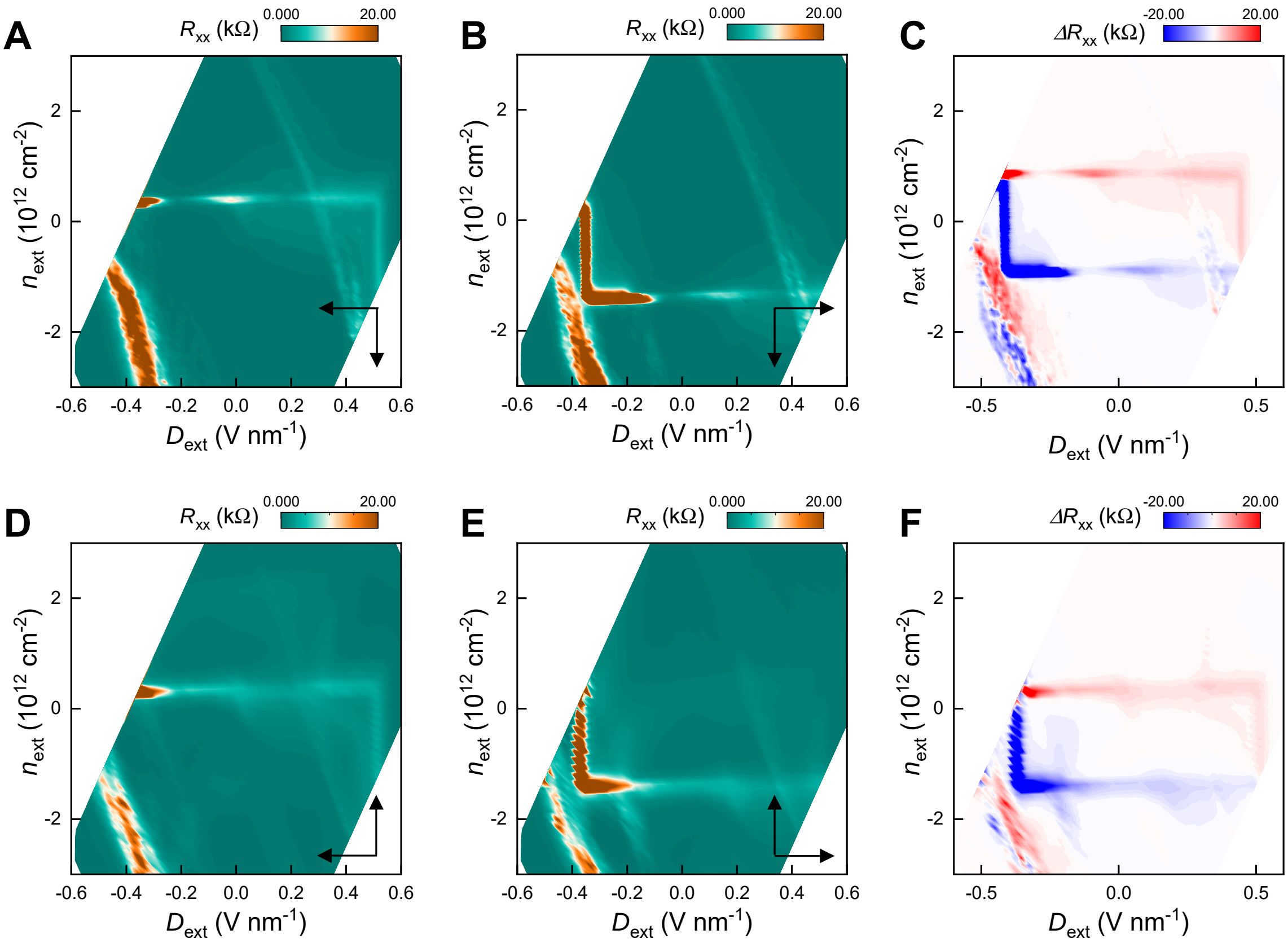

**Fig. S8. Additional data on the transport behaviour of TDBLG device when fast-scanning $D_{ext}$.** (**A-B**) Four-terminal longitudinal resistance $R_{xx}$ *($D_{ext}$, $n_{ext}$)* for backward (A) and forward (B) fast-scans of $D_{ext}$. $n_{ext}$ was step-scanned from electron to hole. (**C**)The resistance difference between (A) and (B)**.** (**D-F**) Same data as (A-C), but with opposite stepping direction of $n_{ext}$. The black arrows indicate the sweeping and stepping directions of $D_{ext}$ and $n_{ext}$.

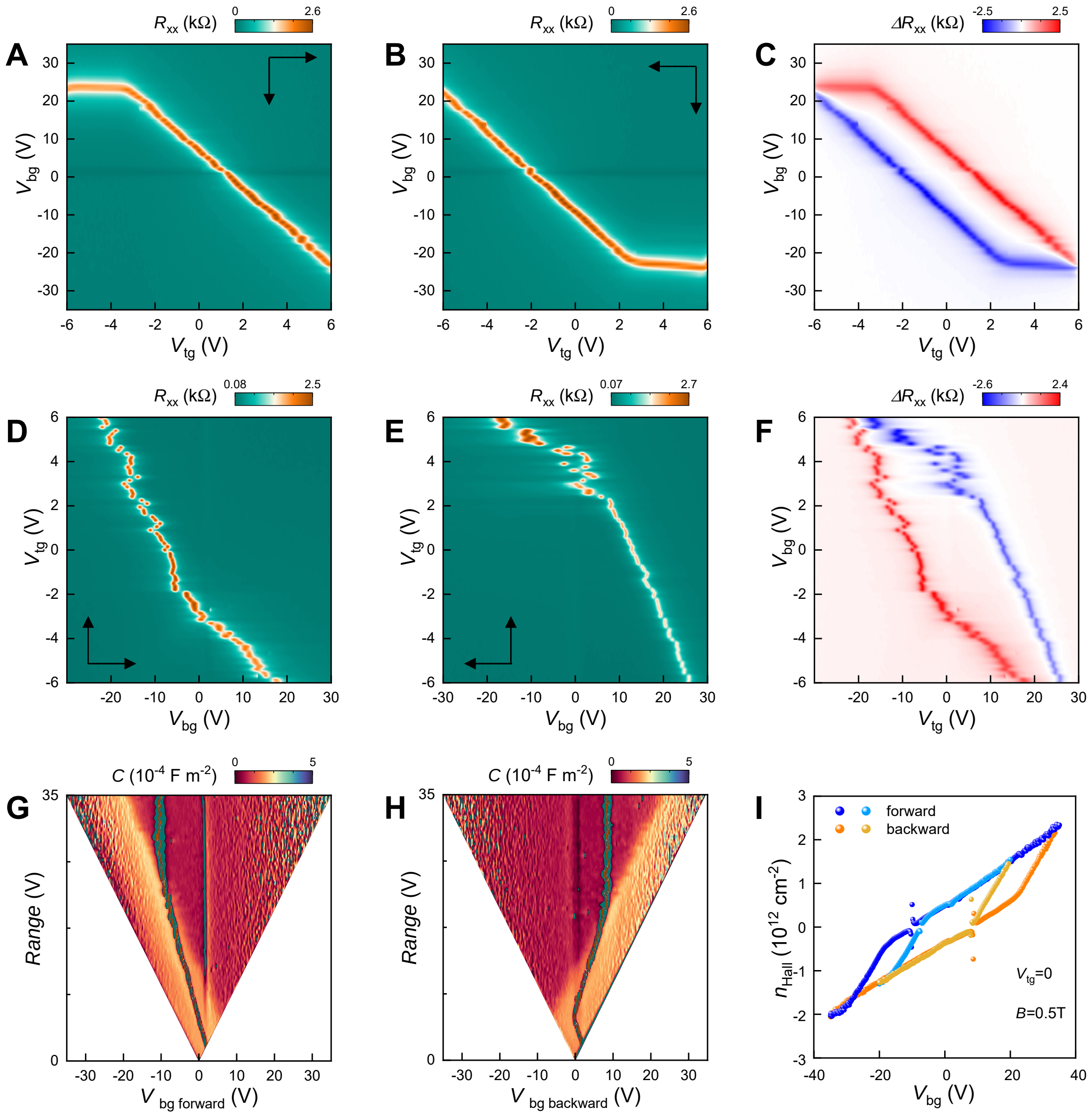

**Fig. S9. More data on the transport properties of moiréless MLG device.** (**A-C**) Dual-gate mapping of four-terminal longitudinal resistance $R_{xx}$ $(V_{tg}, V_{bg})$ for forward (A) and backward (B) fast-scans of $V_{tg}$. $V_{bg}$ was step-scanned from positive to negative. The resistance difference between (A) and (B) is plot in (C). (**D-F**) Dual-gate mapping of four-terminal longitudinal resistance $R_{xx}$ $(V_{bg}, V_{tg})$ for forward (D) and backward (E) fast-scans of $V_{bg}$. $V_{tg}$ was step-scanned from negative to positive. The resistance difference between (D) and (E) is plot in (F). The black arrows indicate the sweeping and stepping directions of gate voltages. (**G-H**) The extracted effective back gate capacitance $C_{bg}$ by gradually increasing the sweeping limits of $V_{bg}$ when $V_{tg} = 0$. The Hall effect was measured under $B = \pm 0.5$T, and $R_{xy}$ was anti-symmetrized. Clearly there is a step change in $C_{bg}$ when we reverse sweeping $V_{bg}$. (**I**) Two sets of line trace of $n_{Hall}$ $(V_{bg})$ with the sweeping limits of ±30 V and ±20 V.

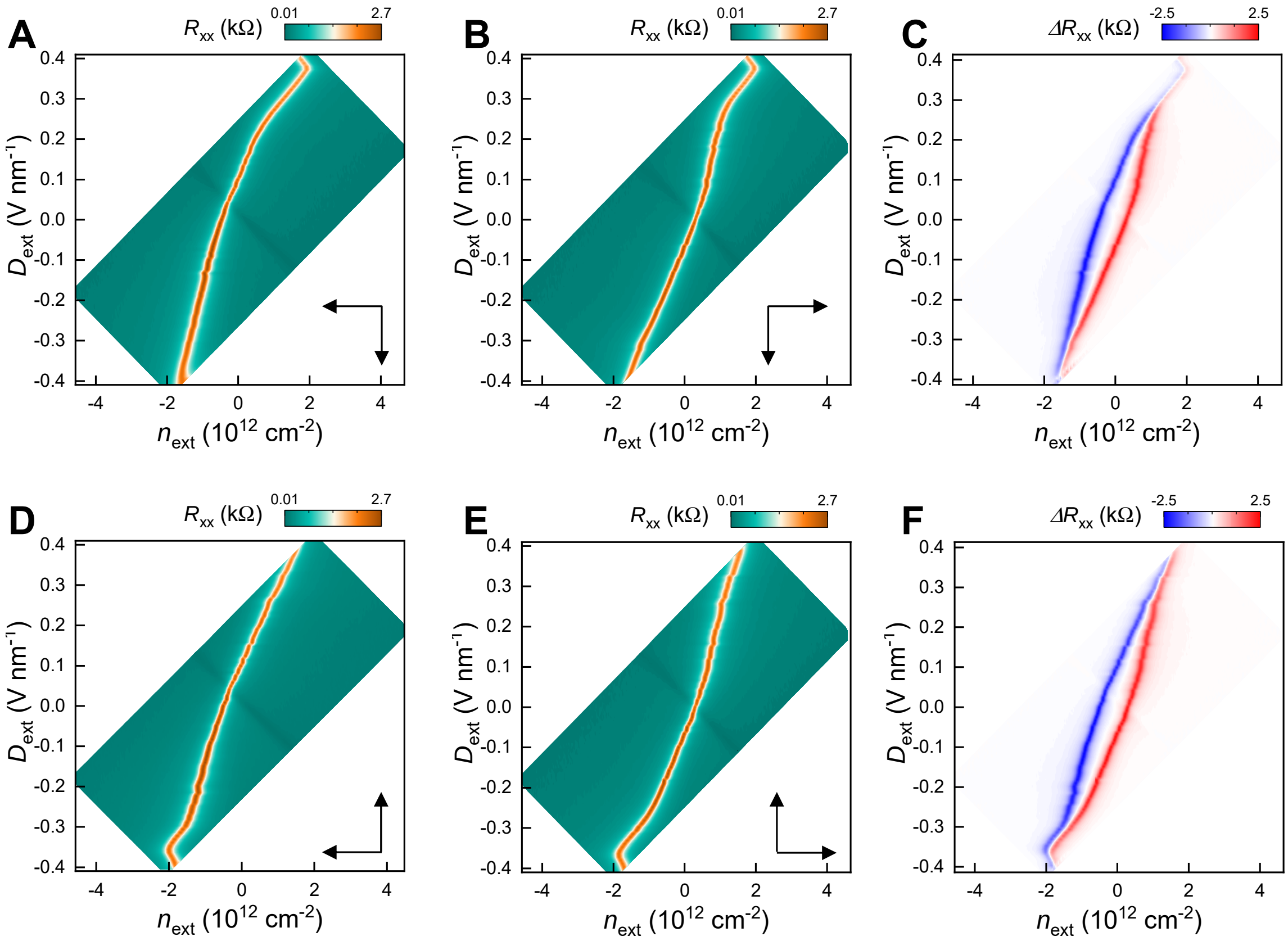

**Fig. S10. Additional data on the transport behaviour of moiréless MLG device when fast-scanning $n_{ext}$.** (**A-B**) Four-terminal longitudinal resistance $R_{xx}$ *($n_{ext}$, $D_{ext}$)* for backward (A) and forward (B) fast-scans of $n_{ext}$. $D_{ext}$ was step-scanned from positive to negative. (**C**) The resistance difference between (A) and (B)**.** (**D-F**) Same data as (A-C), but with opposite stepping direction of $D_{ext}$. The black arrows indicate the sweeping and stepping directions of $n_{ext}$ and $D_{ext}$.

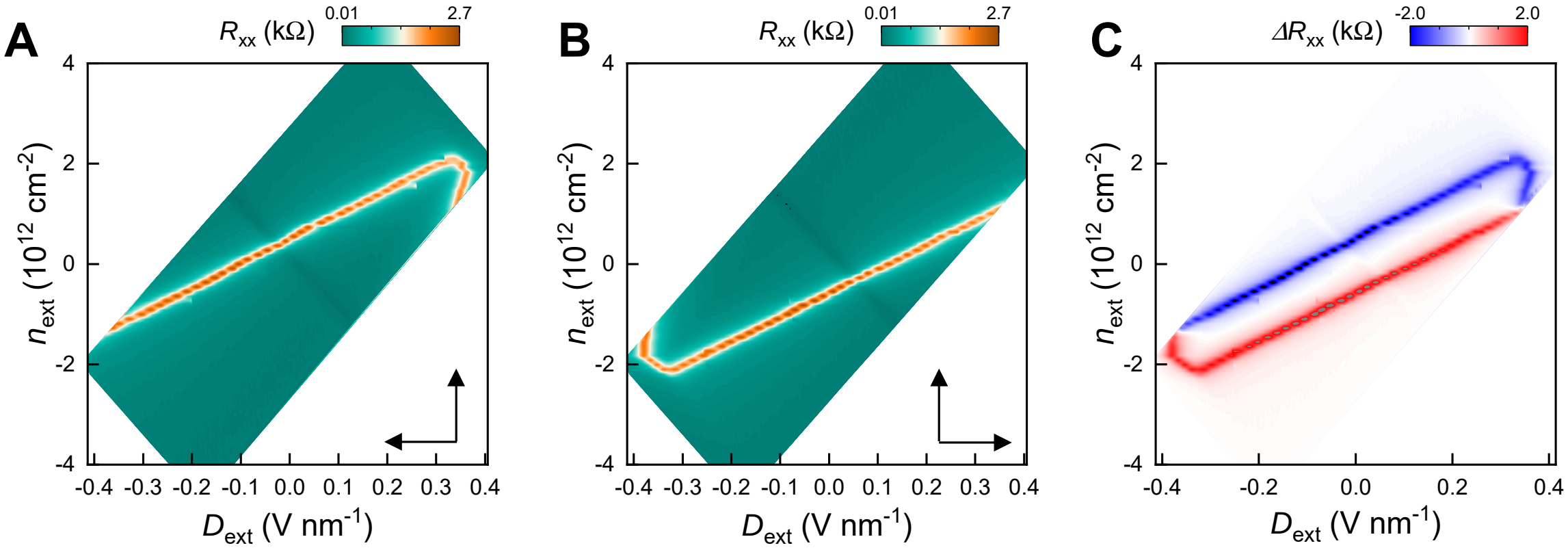

**Fig. S11. Additional data on the transport behaviour of moiréless MLG device when fast-scanning $D_{ext}$.** (**A-B**) Four-terminal longitudinal resistance $R_{xx}$ *($D_{ext}$, $n_{ext}$)* for backward (A) and forward (B) fast-scans of $D_{ext}$. $n_{ext}$ was step-scanned from hole to electron. (**C**) The resistance difference between (A) and (B). The black arrows indicate the sweeping and stepping directions of $D_{ext}$ and $n_{ext}$.

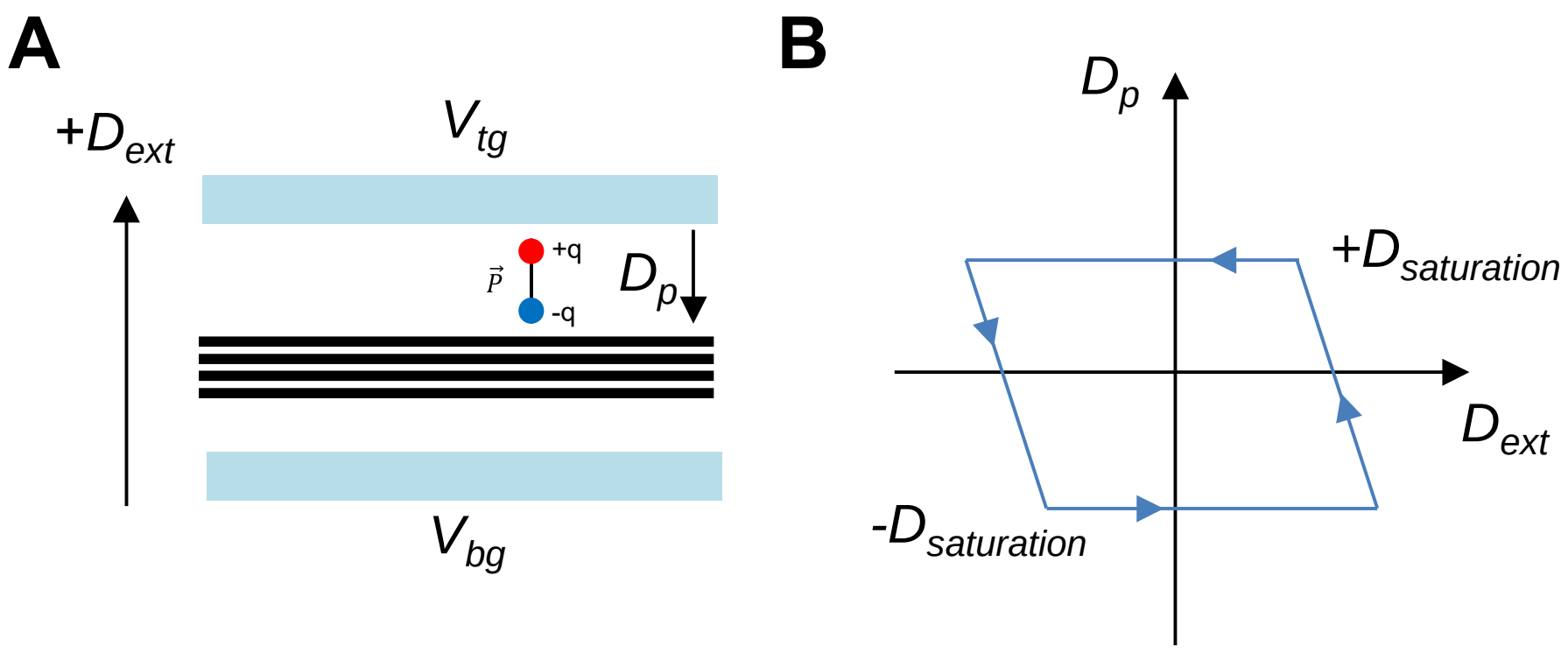

**Fig. S12. Phenomenological scenario with an effective polarization field.** (**A**) Schematic of an electrostatic model. The arrow pointing upward indicates the positive direction of $D_{ext}$. A spontaneous electric polarization $D_p$ is located between graphene and top gate. (**B**) The dependence of the polarization displacement field $D_p$ on external $D_{ext}$. $D_{saturation}$ is the saturation value of $D_p$.